\documentclass[onecollarge,natbib]{svjour2}
\bibpunct{[}{]}{;}{n}{}{,} 
\smartqed  
\usepackage{axodraw}
\usepackage[dvips]{color}
\usepackage{amsmath,amssymb}
\usepackage{epsfig}
\usepackage{graphicx}

\smartqed  

\usepackage{color}
\usepackage{enumerate,array}
\definecolor{navyblue}{rgb}{0.3,0.3,1}
\definecolor{purple}{rgb}{0.6,0,0.5}
\usepackage[colorlinks=true, pdfstartview=FitV,linkcolor=purple, citecolor= navyblue, urlcolor=navyblue]{hyperref}

\newcommand{\be}{\begin{equation}}
\newcommand{\ee}{\end{equation}}
\newcommand{\bea}{\begin{eqnarray}}
\newcommand{\eea}{\end{eqnarray}}
\newcommand{\bno}{\begin{eqnarray*}}
\newcommand{\eno}{\end{eqnarray*}}

\newcommand{\bl}{\begin{large}}
\newcommand{\el}{\end{large}}
\newcommand{\bla}{\begin{Large}}
\newcommand{\ela}{\end{Large}}

\newcommand {\sla} {\slash \hspace*{-2mm}}

\newcommand{\ede}{\end{document}}  

\journalname{Archive of Few-Body Systems} 

\begin{document}

\title{The photon-pion transition form factor: incompatible data or incompatible models? }


\author{J.~P.~B.~C.~de~Melo \and B.~El-Bennich \and T.~Frederico}


\institute{Jo\~ao Pacheco B. C. de Melo (\email{joao.mello@cruzeirodosul.edu.br})
   \at          Laborat\'orio de F\'isica Te\'orica e Computacional, 
                Universidade Cruzeiro do Sul, 01506-000, S\~ao Paulo, Brazil    \\
    \and      Bruno El-Bennich   \at
                Laborat\'orio de F\'isica Te\'orica e 
                Computacional, Universidade Cruzeiro do Sul, 01506-000, S\~ao Paulo, Brazil. \\
    \and     Tobias Frederico \at
                 Instituto Tecnol\'ogico de Aeron\'autica,          12228-900, Sa\~o Jos\'e dos Campos, Brazil.}

\date{Version of \today}

\date{Received: date / Accepted: date}

\maketitle

\begin{abstract}
The elastic and $\gamma \to \pi$ transition form factors of the pion along with its usual static observables are calculated within a
light-front field approach to the constituent quark model.  The focus of this exercise in a simple model is on a unified description 
of all observables with {\em one\/} singly parametrized light-front wave function to detect possible discrepancies in experimental data, 
in particular the contentious large momentum-squared data on the transition factor as reported by BaBar and Belle. 
We also discuss the relation of a small to vanishing pion charge radius with an almost constant pion distribution amplitude and compare 
our results with those obtained in a holographic light-front model. 

\keywords{Light Front Field Theory \and Axial Anomaly \and Electromagnetic Form Factors \and Neutral Pion}
\end{abstract}

\section{Plaidoyer for a consistent and uniform analysis \label{intro}}

The experimental findings of the BaBar Collaboration~\cite{Aubert:2009mc} on the $\gamma \to \pi$ transition form factor in the anomaly-driven 
reaction $\gamma^*\gamma \to \pi^0$ stirred some attention in the hadron community. Indeed, while the BaBar data is in agreement with earlier 
experiments on a domain of squared-momentum transfer below $Q^2=-q^2 \lesssim 10$~GeV$^2$~\cite{Behrend:1990sr,Gronberg:1997fj}, the 
data points at larger $Q^2$ values remarkably exceed the prediction of perturbative QCD (pQCD) in the asymptotic limit~\cite{Farrar:1979aw,Lepage:1980fj}. 
In contrast, a more recent measurement by Belle~\cite{Uehara:2012ag} appears to corroborate the pQCD prediction, although one ought to really 
appreciate the meaning of ``asymptotic": namely that the asymptotic parton distribution amplitude (PDA) on the light front, $\phi_\pi^\mathrm{asy.} = 6x(1-x)$,
is not an appropriate description of the meson's internal structure at scales currently available in experiments~\cite{Segovia:2013eca,Cloet:2013tta,Chang:2013pq}.

On the other hand, at the relevant momentum scale of the $\gamma^*\gamma \to \pi^0$ transition~\cite{Aubert:2009mc}, it was shown
\cite{Roberts:2010rn,Stefanis:2011fv,Bakulev:2011rp,Bakulev:2012nh,Brodsky:2011yv,Brodsky:2011xx,ElBennich:2012ij,Dumm:2013zoa} that PDA 
modifications, as advocated by the studies in Refs.~\cite{Radyushkin:2009zg,Dorokhov:2010zz,Kroll:2010bf,Wu:2010zc,Agaev:2010aq}, lead to form 
factors which deviate drastically from its QCD asymptotic form. The resulting distributions, $\phi(x) \neq \phi_\pi^\mathrm{asy.}$, are constant or at least 
non-vanishing $\forall x \in [0,1]$ and characterize an essentially point-like pion~\cite{Roberts:2010rn}. One comes to a similar conclusion within the framework 
of a light-front quark model~\cite{ElBennich:2012ij} and the impact of the form of the PDA on numerical results has been discussed in detail in 
Ref.~\cite{Bakulev:2012nh}. These overly flat distributions cannot be reconciled with nonperturbative studies of the pion's Bethe-Salpeter 
amplitude~\cite{Cloet:2013tta}. 

The process $\gamma^*\gamma \to \pi^0$ is very interesting in its own right. If one wants to describe the entire domain of experimentally explored 
momentum transfer within a unique theoretical framework, it must simultaneously account for the nonperturbative Abelian anomaly and  the functional 
behavior of perturbative QCD. Studies based on perturbation theory and model input which solely aim at higher-order precision in PDA calculations cannot 
catch the full extent of the nonperturbative nature of light hadron bound states. Essential contributions stemming from the dressing of the fermions and 
gauge bosons are lost and can lead to unnatural predictions, such as a double-dip structure of the pion's PDA, a feature neither congruent with one's 
intuition about the Goldstone boson nor with nonperturbative continuum studies~\cite{Chang:2013pq,Maris:1997tm}.

In practical calculations of the pion's static properties and form factors, one is thus left with the choice to use the most advanced contemporary 
nonperturbative QCD tools, for examples lattice-regularized QCD~\cite{Aoki:2008sm,Bietenholz:2010az} and the combined approach of Dyson-Schwinger and Bethe-Salpeter 
equations~\cite{Bashir:2012fs,Cloet:2013jya}, or alternatively use models whose connection to QCD is not straightforward yet with
its guidance allow for successful numerical results and predictions. We here discuss such a QCD-based  model on the light front 
\cite{deMelo:2002yq,deMelo:2003uk,deMelo:2005cy,Bakker:2013cea} and refer the reader to the more recent approach of light-front holography
discussed by Stan~Brodsky in this meeting~\cite{Brodsky:2013dca}.

In Ref.~\cite{ElBennich:2012ij}, we investigated the effect of varying wave functions in a given established light-front quark model for 
the pion~\cite{ElBennich:2008qa,Mello:2012jv,daSilva:2012gf} applied to the triangle diagram which describes the  $\gamma^*\gamma \to \pi^0$
transition. The nonperturbative contributions to this reaction are not dynamically generated in this approach but encoded in the 
parametrized wave functions and constant mass function of the dressed quark propagators. The point of the exercise is not to test
the state-of-the-art Bethe-Salpeter amplitude for a given decay or transition, but rather to apply this model to all relevant observables. These are 
the static properties, the pion decay constant and electric charge radius, and the elastic and transition form factors. The weak decay constant, 
$f_\pi$, serves to adjust the unique parameter that enters the bound-state wave function and introduces a mass scale. This fixed parameter is 
then used to compute the charge radius and the form factors. No adjustment is made in the process to accommodate a specific observable
unless it is made {\em consistently and uniformly\/} to all observables, the reason for which is rather simple: should we find a wave function, or 
implicitly the pion's PDA since $\Phi_\pi \equiv \Phi\, (\mathbf{k_\perp},x)$, whose functional behavior leads to the rise of the $\gamma \to \pi$ 
transition form factor observed in the BaBar data, the same wave function must yield an elastic form factor as well as a decay constant and
charge radius consistent with well known experimental values~\cite{Beringer:1900zz}. 

As we show in Section~\ref{results}, the light-front calculations with a given wave function reproduce rather well the weak decay constant,
charge radius, elastic form factor and the pion-photon transition form factor {\em if\/} these form factors tend toward the asymptotic pQCD 
limit. It is not possible to reconcile both the BaBar and Belle data above $q^2 \simeq 20$~GeV$^2$ with our model which prefers the Belle
results. Nonetheless, we can adapt the parameter and achieve a transition form factor which partially accounts for the rising tendency of the 
BaBar data. This, however, is at the cost of a rather small charge radius, $\langle \sqrt{r^2_\pi}\, \rangle < 0.4$~fm and a too hard elastic
form factor. Hence, the question arises whether our light-front model is too simple, as other approaches seem to find a compromise between both 
data sets (however without presenting the corresponding values for the weak decay constant, charge radius and elastic form factor of the pion). 
Or is there an inherent tension in the experimental data which are incompatible?

\section{Electromagnetic current and form factors}

In the following, we briefly summarize the theoretical set-up~\cite{ElBennich:2012ij}: kinematics, impulse approximation, and form factor definitions.  
Let us first remind that the matrix element of the electromagnetic current is given in the impulse approximation by the three-point function
\cite{deMelo:1997cb,deMelo:2003uk,ElBennich:2008qa},
\begin{eqnarray}
\hspace{-4mm}
\left\langle p^{\prime }\left\vert J^q_\mu  \right\vert p\right\rangle =     \frac{N_c}{( 2\pi  )^4 }
\int \! d^4 k\, \mathrm{Tr} \left[  \Lambda_{\pi^\prime } ( k,p^\prime  ) S_q( k-p^\prime  ) J^q_\mu (p,p',k )  S_q(k-p)
\Lambda_\pi ( k,p ) S_{\bar q} ( k ) \right ]   +  [ q \leftrightarrow \bar q ] \, , 
\end{eqnarray}
where $N_c=3$ is the color number, $ J^q_{\mu } = \bar q \gamma_\mu  q$ is the electromagnetic current\footnote{since in the light-front 
model the quark-mass function is constant, $M(p^2) = M({p'}^2) = M $, and the wave function renormalization is $Z(p^2) = Z({p'}^2) \simeq 1 $, 
the Ball-Chiu ansatz~\cite{BallChiu,Rojas:2013tza} for the dressed quark-photon vertex reduces to the bare form $\gamma_\mu$. In an approach 
where the mass is generated dynamically, this amounts to treating the light quarks as ``heavy constituent quarks" whose mass is constant for all 
values of $p^2$~\cite{ElBennich:2009vx}. Nevertheless, the Ward-Takahashi identity is preserved in our formalism.}  
and the dressed quark propagator is $S_q (p)=1/(\sla p- M +i\epsilon)$ with the constant quark mass $M$. The vertex functions, 
$\Lambda_{\pi,\pi'}$, represents the Bethe-Salpeter amplitude which projected onto the light-front hyper-surface yields the  
valence wave function of the pion~\cite{deMelo:2002yq,deMelo:2003uk}.
The elastic form factor is extracted from the electromagnetic current via,
\begin{eqnarray}
  \left \langle\pi( p') \left\vert J_\mu^q \left( q^{2}\right) \right\vert  \pi ( p )\right\rangle = \left( p+p^{\prime}\right )_\mu F_\pi^{\mathrm{em}}(q^2) \ ,
\end{eqnarray}
where $q=p'-p$. The weak decay constant of the pion is defined as, 
\begin{eqnarray}
  \left\langle 0 \left\vert A_\mu (0) \right\vert \pi(p) \right\rangle = \imath \sqrt{2} f_\pi\,  p_\mu \ , 
\end{eqnarray}
whose value is experimentally well determined as $f_\pi = 92.4$~MeV~\cite{Beringer:1900zz}.

Along similar lines, we formulated the $\gamma^*\gamma \to \pi^0$ transition form factor in the light-front quark-model approach
\cite{deMelo:1997cb,deMelo:2002yq,deMelo:2003uk,deMelo:2005cy,ElBennich:2008qa}. The matrix element of the neutral pion decay, 
$\pi^0 \to \gamma\gamma$, driven by the Abelian anomaly is given by one unique $CPT$-invariant Lorentz structure. If one photon is off-shell, 
the same matrix element describes the transition amplitude $\gamma^*\gamma \to \pi^0$,
\begin{equation}
  \langle \gamma (p') | J_\mu^q\, |  \pi^0(p) \rangle =e^2\, \epsilon_{\mu \nu \alpha \beta}\,  \epsilon^\nu\!(p')\, q^\alpha p'^\beta F_{\gamma\pi^0}(q^2) \ ,
\end{equation}
where  $\epsilon^\nu (p')$ is the polarization of the real photon, $p=p'+q$ and $p'^2=0, p^2 = m_\pi^2$. Owing to the bosonic symmetrization of the amplitude, 
the transition amplitude receives two contributions:
\begin{equation}
T_{\mu \nu}(q,p') = t_{\mu \nu} (q,p')+ t_{\mu \nu} (p',q) \ .
\label{tensortot}
\end{equation}
Evaluating the traces in spinor and flavor space~\cite{Itzykson:1980rh}, the tensor $t_{\mu \nu} (q,p')$ is given by,
\begin{equation}
    t_{\mu \nu}=\frac{4}{3} \frac{M^2}{f_{\pi}}  e^2 N_c \, \epsilon_{\mu \nu \alpha \beta}\,  q^\alpha p'^{\beta}\, I(q^2) \ ,
\label{tensor}
\end{equation}
where $I(q^2)$ is the scalar loop integral,
\begin{equation}
   I(q^2)=\int \! \frac{d^4k}{(2 \pi)^4}  \frac{1}{((p' - k)^2 - M^2+\imath \epsilon)} \frac{1}{(k^2 - M^2+\imath \epsilon) ((p - k)^2 - M^2+\imath \epsilon) } \ .
\label{integral}
\end{equation}
In Eq.~(\ref{tensor}), the normalization is a consequence of the quark-meson coupling definition in the light-front model,
which is expressed by the Lagrangian ($\hbar = c =1$)~\cite{Frederico:1992ye,Frederico:1994dx},
\begin{equation}
  \mathcal{L}^{\textrm{int}}_{\pi q}  =  - \imath \, \frac{M}{f_\pi} \vec\pi \cdot \bar q\, \gamma_5  \vec\tau\, q \  ,
\end{equation}
where  $\vec\pi$ and $q$ are, respectively, the pion field and quark wave functions and $\vec \tau$ denotes isospin matrices. One may think of this coupling
as the leading term of the full pseudoscalar Bethe-Salpeter amplitude.

After transformation to light-front variables, $\vec k_{\perp}$, $k^+=k^0+k^3$ and $k^-=k^0-k^3$, and integration over the light-front energy, $k^-$,
the final expression for the pion-to-photon transition form factor reads,
\begin{equation}
  F_{\gamma\pi^0}(q^2)  =  \frac{N_c}{6 \pi^3}\frac{M^2}{f_\pi}  \int \frac{dx\, d^2 K_\perp}{(1-x)} \frac{1}{((\vec{K}+x \vec{q})^2_\perp+M^2)(m_\pi^2-M_0^2)} \ ,
\label{formfac1}
\end{equation}
where $m_\pi$ is the pion mass and the reference frame is chosen such that $q^+=q^-=0$ and the momentum transfer, $q_\perp$, is transversal. 
The free-mass operator, $M_0$, is written in terms of the momentum fraction, $x=k^+/p^+$ $(0 < x < 1)$,  and the relative transverse  
$\bar qq$ momentum, $\vec{K}_\perp=(1-x) \vec{k}_\perp-x (\vec{p}-\vec{k})_\perp$,  as:
\begin{equation}
  M_0^2(K_\perp^2,x) =\frac{K_\perp^2+M^2}{x(1-x)} \ .
 \label{freemass}
\end{equation}
Moreover, in the soft (chiral) pion limit, the transition form factor becomes~\cite{Itzykson:1980rh}:
\begin{equation}
   F_{\gamma\pi^0}(0)=\frac{1}{4 \pi^2 f_\pi}
\end{equation}
One can also define a charge radius as the derivative of the transition form factor:
\begin{equation}
    r_{\pi^0}^2 =  6 \, \frac{d F_{\gamma\pi^0}(q^2)}{dq^2}\Big |_{q^2=0} \ .
\label{neutralradius}
\end{equation}

In following Refs.~\cite{Frederico:1992ye,deMelo:1997cb}, we  identify an asymptotic pion wave function in Eq.~\eqref{formfac1}
and introduce the following wave-function ansatz,
\begin{equation}
   \frac{1}{ -m_\pi^2+M_0^2}\   \longrightarrow \ \frac{\pi^{\frac{3}{2}}   f_\pi}{M \sqrt{M_0 N_c}}\, \Phi_\pi(K^2)  \ ,
 \label{asymptoticwave}
\end{equation}
to mimic the soft QCD behavior at low momentum transfer and hard perturbative effects for large $q^2$. The normalization of the 
pion wave function is:
\begin{equation}
   \int d^3 K\  \Phi^2_\pi(K^2) = 1   \ .
\end{equation}
Eventually, we arrive at the following expression for the $\gamma \to \pi^0$ form factor on the light front:
\begin{equation}
   F_{\gamma\pi^0}(q^2) =   \frac{\sqrt{N_c} M}{6 \pi^{\frac{3}{2}}}\int\! \frac{dx\,d^2K_\perp}{x(1-x) \sqrt{M_0}} 
    \frac{\Phi_\pi(K^2)}{(\vec{K} - x \vec{q})^2_\perp+M^2} \ .
\label{formfac}
\end{equation}
In the asymptotic limit, $Q^2=-q^2~\rightarrow~\infty$, pQCD predicts that $Q^2 F_{\gamma \pi^0}= 2f_\pi$~\cite{Farrar:1979aw,Lepage:1980fj},
which we reproduce numerically with Eq.~(\ref{formfac}).

\section{Results and conclusive remarks \label{results}}

Two models of the pion bound-state wave function in Eq.~(\ref{formfac}) are considered: a Gaussian and hydrogen-atom model~\cite{ElBennich:2012ij},
which both depend on a scale parameter $r_\mathrm{nr}$. Including the constituent quark mass, $M$, we thus have two parameters. For the quark
mass we employ common values in the light-front model of Ref.~\cite{deMelo:1997cb,deMelo:2002yq,deMelo:2003uk,deMelo:2005cy,ElBennich:2008qa}, 
whereas $r_\mathrm{nr}$ is fixed by fitting the weak decay constant~\cite{deMelo:1997cb,Beringer:1900zz}: The explicit expressions are written as,
\begin{eqnarray}
    \Phi_\pi (K^2(\vec K_\perp;x))  & = &   \mathcal{N}_\pi \exp \left [ -\frac{4}{3}\, r_{\mathrm{nr}}^2 K^2 \right ] \ , 
 \label{gauss} \nonumber \\  
     \Phi_\pi (K^2(\vec K_\perp;x)) & = &     \frac{\mathcal{N}_\pi}{\big [ r_{\mathrm{nr}}^2+K^2 \big ]^2} \ , 
 \label{h2o}
\end{eqnarray}
with $K^2 (\vec K_\perp ; x) = M_0^2/4 - M^2$ and where $\mathcal{N}_\pi$ is the normalization of the wave function. 

\begin{table}[t]
\setlength{\extrarowheight}{3pt}
\caption{The model's  length scale parameter, $r_{\mathrm{nr}}$, as a function of the constituent quark mass and for $f_\pi=92.4$~MeV. 
The corresponding charge radii are listed next to $r_{\mathrm{nr}}$ for both the neutral and charged pion.}
\begin{center}
\begin{tabular}{ccccc} 
\hline
\noalign{\smallskip}
 {\bf Model}  & $m_{u,d}~[\mathrm{GeV}]$  & $r_\mathrm{nr}~[\mathrm{fm}]$ & $<r_\pi^2>^{\!1/2}~[\mathrm{fm}]$   &  $ <r_{\pi^0}^2>^{\!1/2}~[\mathrm{fm}]$ \\
 \noalign{\smallskip}
\hline \hline 
\noalign{\smallskip}
{ \bf Gaussian}   & 0.220  & 0.345 & 0.637             & 0.683  \\ 
                          & 0.330  & 0.472 & 0.655             & 0.552 \\
{ \bf Hydrogen}   & 0.220  & 0.593 & 0.795             & 0.782 \\
                          & 0.330  & 0.708 & 0.807             & 0.582 \\ 
{ \bf Experiment }   &        &       &  0.672$\pm$0.008~\cite{Beringer:1900zz}  & \\
\noalign{\smallskip}
\hline 
\end{tabular}
\end{center}
\label{tab1}
\end{table}

In Table~\ref{tab1}, we list a range of model parameters which describe reasonable well the charge radius of the pion and reproduce exactly 
the experimental value for the weak decay constant. The corresponding $F_{\gamma\pi} (q^2)$ form factor are plotted for the two models 
and masses in the left panel of Figure~\ref{decays}. As can be seen, with a constituent-mass value $M=330$~MeV, the form factor considerably 
underestimates the experimental data. This is consistent with application of the model to the elastic form factor for which $M=220$~MeV
also provides the best description of the data. Moreover, while the hydrogen model accommodates the BaBar data above 10~GeV$^2$, its 
hardness below this value is incompatible with both the BaBar, Belle and earlier CLEO data. As discussed in Ref.~\cite{ElBennich:2012ij},
the Gaussian model with the parameters $M=220$~MeV and $r_\mathrm{nr}= 0.345$~fm provides overall the most satisfying description of
the pion's static observables and form factors; in Figure~\ref{decays} this wave function corresponds to the solid (black) curve. 

For comparison, we plot in the right panel of Figure~\ref{decays} the transition form factor for both models and $M=220$~MeV along 
with empirical fits, for which we use BaBar's parametric expression for their data~\cite{Aubert:2009mc}, its application to the Belle 
data~\cite{Uehara:2012ag}, as well as Belle's own parametrization of their data. They follow below in this order: \medskip

\begin{figure}[t!]
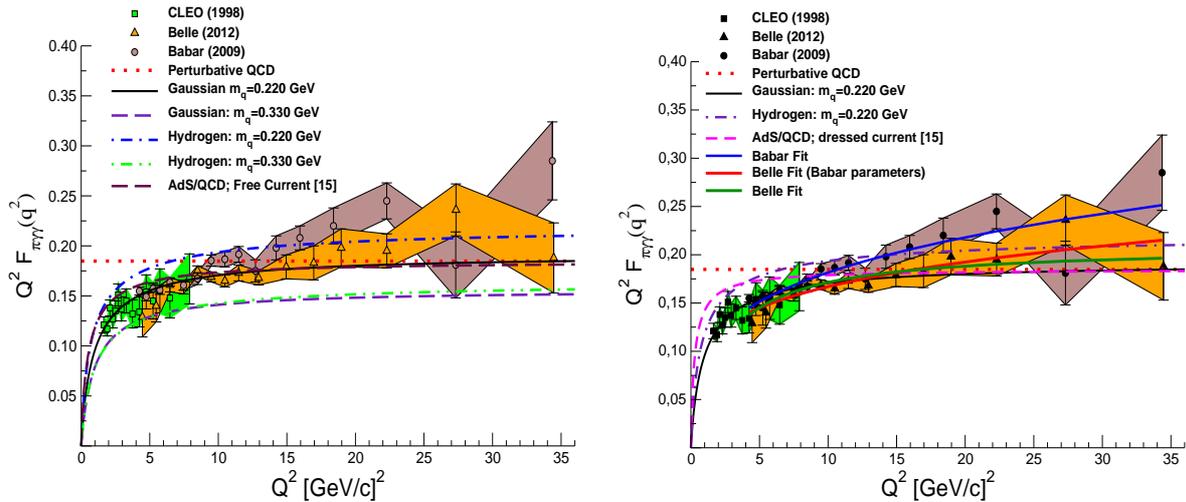

\vspace*{2mm}
\centerline{
\epsfig{figure=lc13fig7.eps,width=7.5cm,height=6.6cm} 
\hspace{0.3cm} 
\epsfig{figure=lc13fig8.eps,width=7.5cm,height=6.6cm} 
}
\caption{The momentum weighted $F_{\gamma\pi}$ transition form factor. In the left panel we compare the light-front model for two sets
of wave functions and masses with experimental data~\cite{Aubert:2009mc,Gronberg:1997fj,Uehara:2012ag} and the AdS/QCD model
expression in Eq.~(\ref{beq30}). In the right panel, the light-front wave-function model which describes most sucessfully {\em all\/} pion 
observables is plotted along with the AdS/QCD-model transition form factor in Eq.~(\ref{beq35}) and the BaBar and Belle fits (see text below).}
\label{decays}
\end{figure}

BaBar: \ $ Q^2 |F(Q^2)|\ = \ A~(\frac{Q^2}{10~\mathrm{GeV}^2})^{\beta} \ \Longrightarrow$ \  
 $\left\{ \begin{array}{ll}
     A = 0.182 \pm 0.002~\mathrm{GeV}  \\
     \beta = 0.250 \pm 0.02~\mathrm{GeV}  
   \end{array}  \right.$

Belle: \ $ Q^2 |F(Q^2)|\ =  \ A_{1}~(\frac{Q^2}{10~\mathrm{GeV}^2})^{\beta_1} \ \Longrightarrow$ \ 
$\left\{ \begin{array}{ll}
  A_1 =  0.167 \pm 0.0036~\mathrm{GeV} \\
  \beta_1 = 0.204 \pm 0.033   
\end{array}  \right.$

Belle: \ $ Q^2 |F(Q^2)|\ =  \ \frac{B~Q^2}{Q^2+C} \ \Longrightarrow$ \ 
$\left\{ \begin{array}{ll}
  B = 0.209~\pm 0.016~\mathrm{GeV}  \\
  C = 2.2~\pm 0.8~\mathrm{GeV}^2 
\end{array} \right. $
\medskip 

As mentioned before, for momentum-squared values above above 10~GeV$^2$, the hydrogen model is most apt at providing 
a reasonable description of the BaBar measurement. The functional form of  $F_{\gamma\pi} (q^2)$ obtained in this domain 
(e.g., the dash-dotted indigo curve in the right panel of Figure~\ref{decays}) is compatible above $Q^2 > 10$~GeV$^2$ with the 
application of the BaBar parametrization to both data sets, namely the blue and red solid lines.  

In Figure~\ref{decays}, we also compare our light-front model with an application of light-front holography following 
Brodsky {\em et al\/}.~\cite{Brodsky:2011xx}, which is based on the AdS/CFT duality. The transition form factor computed in
light-front holography is given by,
\begin{equation}
 Q^2 F_{\gamma\pi^0}(Q^2) =   \frac{4}{\sqrt{3}} \int_0^1\! dx\, \frac{\phi_\pi(x)}{1-x} 
 \left[   1- \exp\left(  -\frac{(1-x) P_{\bar q q} Q^2}{4 \pi^2 f^2_\pi\, x } \right)  \right] \ ,
\label{beq30}
\end{equation}
where $\phi(x)=\sqrt{3}f_{\pi} x(1-x)$ is the asymptotic pion distribution function and $P_{q\bar{q}}$ is the probability 
to find the valence $\bar qq$ state. The value $P_{\bar q q}=1$ is consistent with the leading-order pQCD result~\cite{Lepage:1980fj}. 
In case of a dressed current (see Eq.~(35) in Ref.~\cite{Brodsky:2011xx}), which effectively corresponds to a superposition of Fock 
states, and using a twist-2 distribution function for $\Phi_\pi(x)$, the form factor is modified to,
\begin{equation}
   Q^2 F_{\gamma\pi^0}(Q^2)=  8 f_{\pi} \int_0^1\! dx\, \frac{1-x}{(1+x)^3}   \left [ 1-  x^{ Q^2 P_{\bar qq}/(8  \pi^2 f^2_\pi)  }  \right]\ ,
\label{beq35}
\end{equation}
which is represented by the dashed magenta curve in the right panel of Figure~\ref{decays}.

\begin{figure}[t!]
\vspace*{1mm}
\centerline{
\epsfig{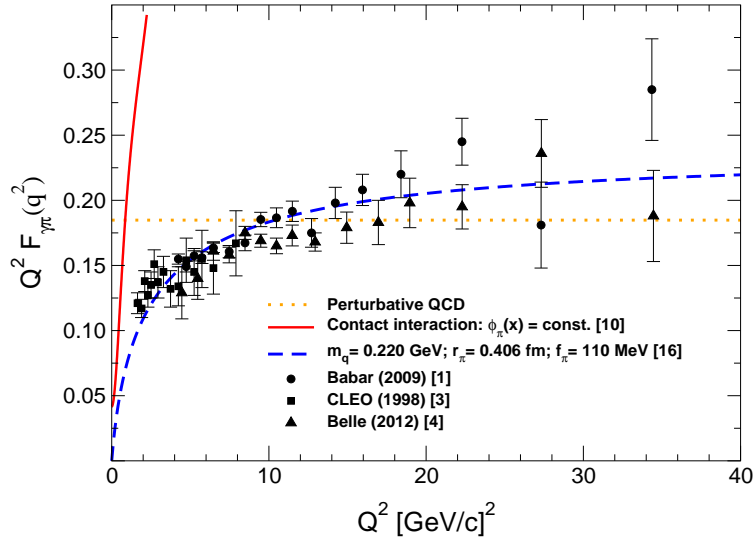} 
}
\caption{The effect of a decreasing pion charge radius is demonstrated with the blue dashed curve: the Gaussian wave function ansatz
in the present light-front model is adjusted so that the functional form of $F_{\gamma\pi }(Q^2)$ is in mutual agreement with the BaBar and
Belle data sets for $Q^2 > 10$~GeV$^2$ and likewise reproduces the data for lower four-momentum transfer. 
As a consequence, the charge radius decreases considerably while the weak decay constant increases. Taken to an extreme, a 
self-consistently regularized and symmetry-preserving contact interaction model~\cite{Roberts:2010rn} describes the pion as a point-like 
particle with, $\phi_\pi (x) = C\,  \ \forall x \in [0,1]$, and yields the transition form factor plotted as a red solid curve. }
\label{contact}
\vspace*{-1mm}
\end{figure}

The two form factor expressions obtained in the AdS/QCD models reproduce the asymptotic pQCD prediction~\cite{Lepage:1980fj}, 
$Q^2 F_{\gamma \pi }=2 f_{\pi}$, and are in agreement with the Belle data at larger $q^2$ and our Gaussian wave function ansatz 
with $M=220$~MeV (black solid curve). They are thus in contradiction with the Babar data above, say, 20~GeV$^2$.

We close this presentation with a last comparison in Figure~\ref{contact}. In there, we adapt the wave function parameter $r_\mathrm{nr}$ of
the Gaussian wave function ansatz which best simultaneously and self-consistently reproduces the static pion and form factor data (black solid 
curve in Figure~\ref{decays}). This leads to a transition form factor which partially accounts for the rising tendency of the BaBar data, yet yields 
a rather small charge radius, $\langle \sqrt{r^2_\pi}\, \rangle \simeq 0.4$~fm. Moreover, the elastic form factor using this modified wave function 
is too hard and in conflict with all experimental data~\cite{ElBennich:2012ij}. This is qualitatively in agreement with the result obtained by means 
of a vector-vector contact interaction applied to the Bethe-Salpeter and gap equations in rainbow-ladder truncation. Since this interaction leads 
to quadratic divergences, it must be treated with a symmetry-preserving regularization~\cite{Roberts:2010rn}. The consequences of this 
interaction are a constant constituent-like quark mass and a point-like pion with a flat distribution amplitude, $\phi_\pi(x) =C$, which implies 
that all form factors asymptotically approach a constant. This behavior is observed in the red solid curve in Figure~\ref{contact} which 
corresponds to the self-consistent rainbow-ladder treatment of the contact interaction.

\begin{acknowledgements}
This work was supported by the S\~ao Paulo State Research Foundation (FAPESP) under grants nos.~2009/00069-5, 2009/53351-0, 2009/51296-1 
and  2012/17396-1. T.~Frederico also acknowledges support by the  National Council for Scientific and Technological Development  (CNPq).
J.~P.~B.~C.~de Melo  and T.~Frederico thank Nicos Stefanis for the LC2013 organization and the invitation.  B.~El-Bennich appreciated helpful 
communication with Adnan~Bashir and is grateful to Laura Xiomara Guti\'errez-Guerrero for providing the transition form factor data in Figure~\ref{contact}.
\end{acknowledgements}

\end{document}